\documentclass[preprint]{emulateapj} 

\shorttitle{Thermal Evolution of EGP Radii} 
\shortauthors{Batygin Stevenson Bodenheimer}

\begin{document}
 
\title{Evolution of Ohmically Heated Hot Jupiters}  
\author{Konstantin Batygin$^1$, David J. Stevenson$^1$ \& Peter H. Bodenheimer$^2$} 

\affil{$^1$Division of Geological and Planetary Sciences, California Institute of Technology, Pasadena, CA 91125} 
\affil{$^2$UCO/Lick Observatory, University of California, Santa Cruz, CA 95064 }

\email{kbatygin@gps.caltech.edu}

\begin{abstract}
We present calculations of thermal evolution of Hot Jupiters with various masses and effective temperatures under Ohmic dissipation. The resulting evolutionary sequences show a clear tendency towards inflated radii for effective temperatures that give rise to significant ionization of  alkali metals in the atmosphere, compatible with the trend of the data. The degree of inflation shows that Ohmic dissipation, along with the likely variability in heavy element content can account for all of the currently detected radius anomalies. Furthermore, we find that in absence of a massive core, low-mass hot Jupiters can over-flow their Roche-lobes and evaporate on Gyr time-scales, possibly leaving behind small rocky cores.
\end{abstract}

\section{Introduction}
Over the last decade, novel discoveries of transiting extra-solar planets have often been fraught by unexplained radius anomalies. In particular, many gas giant planets, residing on orbits in extreme proximity to their host stars, have been found to have much larger radii than what was previously thought possible in the context of ``standard" gas giant theory \citep{1982AREPS..10..257S,1999Sci...286...72G}. A number of possible explanations to this problem have been proposed, most notably tidal heating \citep{2001ApJ...548..466B,2003ApJ...592..555B}, kinetic heating due to breaking gravity waves \citep{2002A&A...385..156G}, enhanced opacity \citep{2007ApJ...661..502B}, semi-convection \citep{2007ApJ...661L..81C} and turbulent burial of heat by a mechanical greenhouse effect \citep{2010ApJ...721.1113Y}. However, it appears unlikely that any of the above solutions can simultaneously explain all of the observed anomalies \citep{2010SSRv..152..423F}. Recently, a new Ohmic heating mechanism, that relies on the electro-magnetic interactions between atmospheric flows and the planetary magnetic field, has been suggested as a promising explanation to the inflation problem \citep{2010ApJ...714L.238B}. 

A characteristic that is unique to hot Jupiters is the enormous amount of incident energy that they receive from their host stars. The extreme irradiation naturally results in high atmospheric temperatures (sometimes in excess of 2000K) \citep{2009ApJ...699.1487S}, and  the latitudinal dependence of this heating leads to fast, super-rotating equatorial jets with characteristic wind-speeds of order $\sim$ 1 km/s \citep{2009ApJ...699..564S}. Consequently, the atmospheres of hot Jupiters are hot enough to thermally ionize alkali metals, that are present in trace abundances, and give rise to a finite electrical conductivity of order $\sim 10^{-3} - 1$ S/m, depending primarily on the effective temperature. Rapid advection of the ions by the jets, while immersed in the planetary magnetic field, leads to an induction of an electro-motive force that sets up electrical current loops through the deep interior and the atmosphere of the planet. These electrical currents give rise to Ohmic heating as they flow throughout the planet. 

It has been shown, in the context of a static structural model, that the level of resulting dissipation in the interior is approximately that required to maintain the radii of hot Jupiters \citep{2010ApJ...714L.238B}. However, time-evolution of Ohmically heated gas-giants remains an open question, since the degree of dissipation exhibits strong dependence on the structure of the planet, and particularly on the planet's $T_{eff}$, mass, and the location of the radiative/convective boundary. More explicitly, the thermal structure can be affected by the heating and its radial variation as well as the elapsed time. At the same time, the conductivity arises from thermal ionization, causing the interior structure to feedback on the heating. Thus, the structural profile of the planet dictates the Ohmic heating distribution. As a result, self-consistent evolutionary calculations are required to assess whether Ohmic dissipation can indeed resolve the inflation problem. To perform such calculations is the primary purpose of this study.

Here, we construct an approximate model with specific assumptions, that couples calculations of Ohmic heating and structural evolution, and show how this heating together with the likely variability of heavy element abundances can provide an explanation for all of the currently observed radius anomalies. We compute the maximum radii attainable by evolved Hot Jupiters as functions of planetary mass and effective temperature and find a tendency for larger radii for effective temperatures that correspond to the onset of significant conductivity arising from alkali metals in the atmosphere. This behavior is compatible with the trend of the data. Furthermore, our calculations suggest that in absence of a relatively massive, high-metallicity core, low-mass hot Jupiters can expand beyond their Roche-lobes and spill their envelopes onto their parent stars on billion year time-scales, possibly leaving behind small rocky cores.

The paper is structured as follows. In section 2, we consider the energetics of the Ohmic mechanism, and show that in steady state, the degree of dissipation is limited by the thermodynamic efficiency of the atmosphere. In section 3, we present a simple model for the damping of the global circulation by the Lorentz force, and show that characteristic efficiency of the Ohmic mechanism is of order a few percent. In section 4, we describe the coupled Ohmic heating/themal evolution model that we employ in our calculations. In section 5, we present the theoretical curves that delineate mass-radius-$T_{eff}$ space and compare our results with observational data. We conclude and discuss avenues for future improvement of the Ohmic inflation model in section 6.

\section{Work - Ohmic Dissipation Theorem}

Let us begin by considering the energetics of the Ohmic dissipation mechanism. We start by writing out the invicid Navier-Stokes equation, where turbulent stresses have been neglected and the last term is the Lorentz force:
\begin{equation}
\frac{D \vec{v}}{Dt} = -2 \vec{\Omega} \times \vec{v} - \frac{\vec{\nabla}P}{\rho} + \vec{g} +  \frac{\vec{J} \times \vec{B}}{\rho}.
\end{equation}
In the above equation, $\vec{v}$ is the velocity,  $\vec{\Omega}$ is the rotational velocity, $P$ is pressure, $\rho$ is density, $g$ is the gravitational acceleration, $\vec{J}$ is the current density and $\vec{B}$ is the background planetary magnetic field. There exists a vast literature on the subject of energy budgets of the various terms in the Navier-Stokes equation (see for example \citet{1992aitd.book.....H} and \citet{1992phcl.book.....P}). Although a discussion of the global picture is important for understanding the details of atmospheric dynamics, the aim of our study is limited to the Ohmic mechanism, so we shall focus our discussion on the Lorentz term. The power, or rate of change of kinetic energy energy of the fluid per unit volume, provided solely by the Lorentz force is
\begin{equation}
\left( \frac{\rho}{2}\frac{D v^2}{Dt} \right)_L= \vec{v}\cdot \vec{J} \times \vec{B}
\end{equation}
If the quantity on the right is positive, the Lorentz force adds kinetic energy to the flow, while if it is negative, kinetic energy is drained from the flow. Recall that MHD Ohm's law reads
\begin{equation}
\vec{J}=\sigma(\vec{v}\times\vec{B}-\nabla\Phi)
\end{equation}
where $\sigma$ is the electrical conductivity and $\Phi$ is the electric scalar potential. Using a standard vector-calculus identity, we can write $\vec{v}\cdot\vec{J}\times\vec{B} = - \vec{J}\cdot\vec{v}\times\vec{B}$. Consequently, the energy equation can be re-written as
\begin{equation}
\vec{v}\cdot \vec{J} \times \vec{B} = -\frac{J^2}{\sigma} - \vec{J} \cdot \nabla \Phi
\end{equation}
The first term on the RHS, is the Ohmic dissipation. Upon integration, by Gauss's theorem, the last term vanishes, since we require no radial current at the outer boundary:
\begin{equation}
\int\int\int \vec{J} \cdot \nabla  \Phi dV = \int\int\int \nabla \cdot ( \vec{J} \Phi) dV = \oint (\vec{J} \Phi) \cdot d\vec{a} = 0
\end{equation}
As a result, we discover that in steady-state, Ohmic dissipation is work done by the flow:
\begin{equation}
\int\int\int  \left( \frac{\rho}{2}\frac{D v^2}{Dt} \right) dV = - \int\int\int    \frac{J^2}{\sigma} dV
\end{equation}
It is important to note, however, that the work done by the flow is limited by the efficiency factor i.e. the fraction of insolation that is available to do useful work. In practice, this means that the total Ohmic dissipation rate should be rather insensitive to the magnetic field strength, once the field is larger than some critical ``saturation" value. Particularly, in the saturated case, the flow velocity should scale inversely with the magnetic field. Thus, in such a regime, changing the field will not change dissipation or the planet radius. Rather the efficiency factor plays the governing role. Although numerical simulations are required to quantitatively understand the saturation field strength accurately, based on dimensionless number analysis, it is likely that the critical field strength is not overwhelmingly high ($\sim 1$ Gauss or so). 

\section{Magnetic Damping of the Global Circulation \\ and the Efficiency of the Ohmic Mechanism}

In the limit where the Lorentz force is dynamically negligible, the wind speeds may be interpreted from global circulation models (GCMs). In this case, the computation of Ohmic dissipation is straight-forward, although it is noteworthy that GCM's carry an intrinsic range of numerical uncertainty \citep{2008ApJ...675..817C, 2010exop.book..471S, 2011MNRAS.tmp..370H}. For sufficiently high atmospheric ionization levels and/or planets with strong magnetic fields, however, the onset of flux-freezing can act to slow down the winds, effectively capping the degree of Ohmic dissipation. The total Ohmic dissipation can be written as some fraction of the absorbed sunlight; we refer to this as the efficiency, $\epsilon$. This efficiency must obviously go to zero as one goes to very low conductivities or low background field strength and is limited from above by thermodynamic (Carnot efficiency) considerations. It is further noteworthy that only a part of the total Ohmic dissipation ($\sim5 $\% for $T_{eff} \gtrsim 1400$K; $\sim 10^{-4} $\% for $T_{eff} = 1000$K) is deposited in the region of relevance to our inflation mechanism (the convective interior); the rest is dissipated at higher levels\footnote{This is because for lower $T_{eff}$ the atmosphere is characterized by higher resistance, and thus dissipates fractionally more energy.} (see Figure 4 and the associated discussion). The primary aim of this section is to derive a quantitative estimate for the Ohmic efficiency.

The limitation to efficiency can arise in two ways: through the reduction of the wind speeds as a direct consequence of the Lorentz force or through the reduction of the equator to pole temperature difference because of meridional circulation. We have not performed full MHD global circulation models. Instead, following \citet{1977JAtS...34..263S} and \citet{1980JAtS...37..515H}, we develop a simple, analytical model for the mid-latitude circulation in hot Jupiters, crudely accounting for magneto-hydrodynamic interactions.

Let us begin by simplifying the Lorentz force. Since the electric field is of the same order of magnitude as $\vec{v} \times {B_{dip}}$ as defined by the boundary condition, we write $\vec{J} \sim \sigma (\vec{v} \times {B_{dip}})$. Collecting the conductivity, magnetic field and fluid density into a damping timescale and retaining only the component that opposes the flow, we have
\begin{equation}
\frac{\vec{J} \times \vec{B}}{\rho} \sim  - \frac{\sigma \vec{v} B^2}{\rho} \sim - \frac{\vec{v}}{\tau_L}
\end{equation}
In other words, the effect of the Lorentz force is to introduce a term into the equation of motion that can be approximated as a Rayleigh drag with characteristic timescale  
\begin{equation}
\tau_L = \rho/\sigma B^2 \sim 10^6 (\frac{\rho}{0.1\textrm{kg/m}^3})  (  \frac{0.1\textrm{S/m}}{\sigma} ) (\frac{10^{-3} \textrm{T}}{B})^2             \textrm{sec.}
\end{equation}
A damping time-scale of the same functional form was used in GCM simulations of \citet{2010ApJ...724..313P} to mimic the the Lorentz force, in order to obtain an estimate of Ohmic dissipation in the atmosphere. As will be revealed below, this term introduces both direct drag as well as meridional circulation into an otherwise geostrophic solution. 

Next we assume steady state and azimuthal symmetry, so all time and zonal derivatives in the Navier-Stokes equation vanish. Finally, under an assumption of a constant Coriolis parameter (f-plane approximation), the horizontal Boussinesq equations of motion reduce to an Ekman balance:
\begin{equation}
f v_y = \frac{v_x}{\tau_L} 
\end{equation}
\begin{equation}
f v_x = -\frac{1}{\rho_0}\frac{\partial P'}{\partial y} - \frac{v_y}{\tau_L} 
\end{equation}
where we have expressed all quantities in a local cartesian coordinate system i.e. $v_{\phi} \rightarrow v_x$, $v_{\theta} \rightarrow v_y$, and $P'$ is the component of pressure that is not compensated by gravity. Under the same assumptions, the vertical momentum equation, neglecting Coriolis effects simplifies to the hydrostatic equation
\begin{equation}
\frac{1}{\rho_0} \frac{\partial P'}{\partial z} = \alpha g \Theta'
\end{equation}
where $\alpha=1/\left<\Theta\right>$ is the coefficient of thermal expansion and $\Theta'$ is the potential temperature departure from the background state $\bar{\Theta}(z)$. Stable stratification is implicit in the problem and we write the heat equation in accord with Newtonian cooling:
\begin{equation}
v_z \left( \frac{d \bar{\Theta}}{d z} \right) = - \frac{\Theta' - \Theta^{rad}}{\tau_N}
\end{equation}
where $\Theta^{rad}$ is the radiative deviation from the background state, and $\tau_N=c_p P / 4g \sigma_{SB} T_{eff}^3$ is the Newtonian cooling timescale. The physical meaning of $\Theta^{rad}$ is that if no meridional flow is present to transport heat from the equator to the pole, the potential temperature profile takes the form $\Theta = \bar{\Theta} + \Theta' = \bar{\Theta} + \Theta^{rad}$. 

\begin{figure*}[t]
\includegraphics[width=1\textwidth]{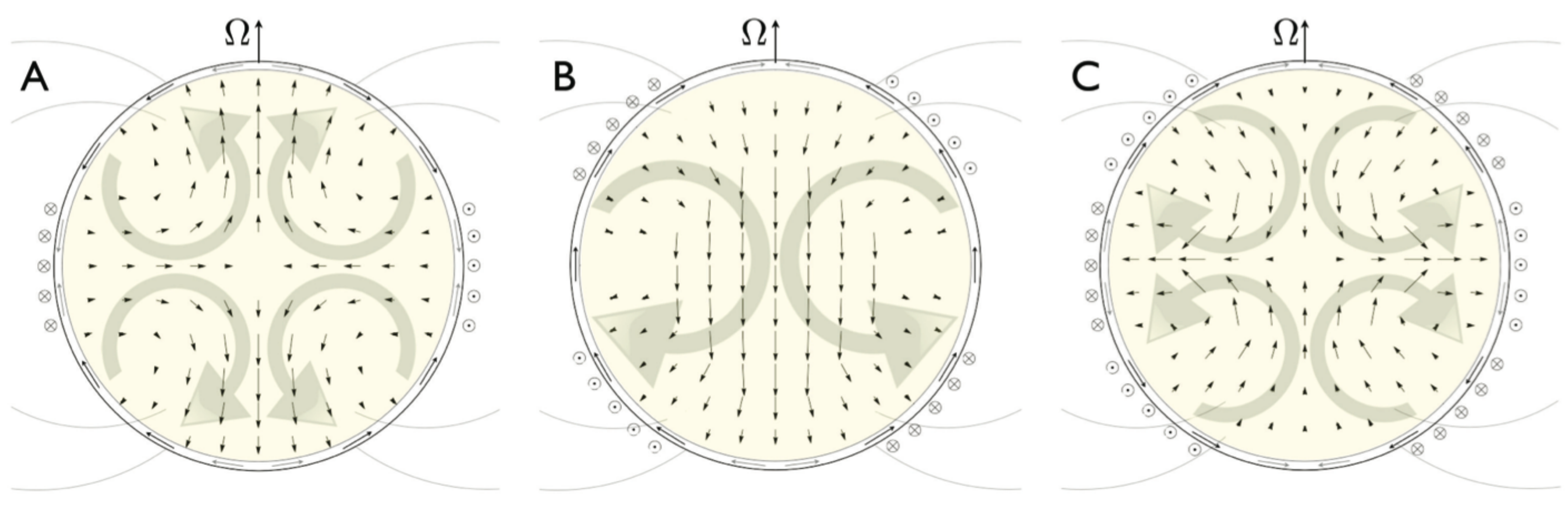}
\caption{A comparison between interior current geometries, induced by (A) a single jet (B) two counter-rotating jets, each in one hemisphere and (C) a triple jet with retrograde equatorial flow. The small arrows inside the planet show the induced current (actual model output). Additionally, dipole magnetic field lines as well as graphical representations of the currents are presented. Atmospheric flows are represented as into-the-board and out-of-the-board arrows on the sides of the planets. } 
\end{figure*}

To obtain a solution for the meridional circulation, let us introduce a stream-function $\Psi(y,z)$ such that $v_y = \partial \Psi/\partial z$ and  $v_z = -\partial \Psi/\partial y$. Upon differentiating the hydrostatic equation with respect to $z$, differentiating the y-momentum equation with respect to $y$ and subtracting the first from the latter, we obtain a modified thermal wind equation. 
\begin{equation}
f \frac{\partial v_x}{\partial z} = -\alpha g \frac{\partial \Theta'}{\partial y} -\frac{1}{\tau_L}\frac{\partial v_y}{\partial z}
\end{equation}
Multiplying through by $d \bar{\Theta}/dz$, expressing the velocities in terms of the stream-function, and applying the heat equation, we arrive at a differential equation for the potential temperature perturbation:
\begin{equation}
\frac{\partial^2 \Theta'}{\partial z^2} + \frac{\tau_N}{\tau_L}\left( \frac{N^2}{f^2} \right)\frac{1}{1+(f\tau_L)^{-2}} \frac{\partial^2 \Theta'}{\partial y^2} = \frac{\partial^2 \Theta^{rad}}{\partial z^2}
\end{equation}
where $N^2 = g \left< \Theta \right>^{-1} d \Theta / dz$ is the square of the Brunt-Vaisala frequency. Note that the Lorentz force is not necessarily anti-parallel to the zonal velocity so the corespodence we use here is designed only to find the effect of the drag. Next we take the solutions to be of the functional form:
\begin{eqnarray}
\Psi &=& \Psi_0 \sin(\pi y / R) \sin(\pi z/H) \nonumber \\
v_x &=& - v_{x0} \sin(\pi y / R) \cos(\pi z/ H) \nonumber \\
\Theta^{rad} &=& \Theta_0^{rad} \cos(\pi y / R) \sin(\pi z/ H) \nonumber \\
\Theta' &=& \Theta_0' \cos(\pi y / R) \sin(\pi z/ H)
\end{eqnarray}
where $H$ is the atmospheric scale-height, $R$ is the planetary radius, and all terms with the subscript "0" are undetermined constants. Upon substitution of these solutions into the above equation, we obtain an algebraic expression for the actual potential temperature deviation as a function of the radiative equilibrium deviation and other system parameters. This expression can be substituted back into the thermal wind equation to yield an order of magnitude equation for zonal wind velocity reduction due to the drag:
\begin{equation}
v_{x} \sim  \frac{g H }{f R} \frac{\Delta \Theta^{rad}}{\left< \Theta \right>}\left( 1 + \left({f \tau_L}\right)^{-2} + \left(\frac{N}{f} \frac{H}{R} \right)^2 \left( \frac{\tau_N}{\tau_L} \right) \right)^{-1}
\end{equation}
where $\Delta \Theta^{rad}$ is the equator-to-pole potential temperature difference that the planet would have if no meridional circulation was present. Consider the limiting cases of this equation. If we take the drag to be negligible ($\tau_L \rightarrow \infty$), we have $\Theta' = \Theta^{rad}$, so the global potential temperature becomes $\Theta = \bar{\Theta} + \Theta^{rad}$, as $v_y \rightarrow 0$. Thus the wind velocity equation reduces to the standard thermal wind equation. If we consider the drag to be the dominant effect ($\tau_L \rightarrow 0$), then the potential temperature perturbation again becomes equivalent to the radiative potential temperature perturbation, since all circulation effectively stops. This is a regime where the above treatment of the Lorentz force is highly inaccurate. The transition between the two regimes takes place when $f\tau_L \sim 1$.  

The functional form for the stream-function that we introduced above satisfies the no-normal flow boundary conditions at the edges of the modeled ``cell". The circulation it implies, however, is somewhat unrealistic and is not globally observed on either terrestrial or solar system gas giant atmospheres. This is largely because in order to obtain the analytical solution presented above, we made a series of simplifying approximations and implicitly omitted any discussion of turbulent Reynolds stresses, $ \nabla \cdot (v_x' v_y')$. In a detailed model for atmospheric circulation, the resulting eddy momentum fluxes are essential to closing the angular momentum cycle on gas giant planets \citep{2010JAtS...67.3652L}, although formally the above solution also closes the angular momentum cycle \citep{1977JAtS...34..263S}. Consequently, we stress that the above picture is not intended to be a quantitatively good representation of the atmospheric dynamics on hot Jupiters. However, we do suggest that the model yields the correct behavior of the wind velocity reduction, to an order of magnitude. 

Direct application of the above formalism to our models is not possible because of the large variation in conductivity and density. However, we can still use the resulting quantitative estimate of Ohmic efficiency as a guide, since any reasonable choices of parameters (e.g. the scale factors in equation 8), yield a Lorentz damping timescale, $\tau_{L}$, that is far longer than the rotational timescale, implying a limited effect on the zonal winds. In other words, for most hot Jupiter regimes, it is reasonable to approximate the effect of the Lorentz force as a Rayleigh drag. The efficiency of conversion of zonal wind energy into heating is then given by the ratio of the kinetic energy dissipated by the drag term in the equations of motion to the incident stellar flux:
\begin{eqnarray}
\epsilon &\sim&  \rho v_{\phi}^2 H / (\tau_L \sigma_{SB} T_{eff}^4) \nonumber \\
&\sim& 0.01 (\frac{\rho}{0.1\textrm{kg/m}^3}) (\frac{v_{\phi}}{1\textrm{km/s}})^2 (\frac{H}{1000\textrm{km}}) (\frac{1500\textrm{K}}{T_{eff}})^4
\end{eqnarray}
A value of this order is in accord with the results of simplified GCM simulations of hot Jupiters where magnetic effects are modeled as drag \citep{2010ApJ...724..313P}. In order to qualitatively explore the effect of variable efficiency, we compute three sets of models, characterized by 1\%, 3\%, and 5\% Ohmic efficiencies. Note that at low ($T_{eff} \sim 1000K$) temperatures, such efficiencies are likely to be overestimates, because of very low conductivity. However, the mechanism at very low conductivity is inactive anyway, so we retain the same efficiencies for consistency. From a computational point of view, a constant efficiency means that although the dissipation is a function of the radial coordinate, it is scaled such that the integrated Ohmic heating amounts to 1\%, 3\%, or 5\% of the insolation. In other words, we only adjust the magnitude of the heating, not its radial distibution. 

\section{Coupled Ohmic Heating/Structural Evolution Model}

In this study, the calculation of the Ohmic heating was performed in effectively the same way as in \citep{2010ApJ...714L.238B}, but with the atmospheric temperature structure computed rather than assumed. In other words, as the planet evolves along its evolutionary sequence, the conductivity profile everywhere in the planet is recalculated at every time-step. Naturally, the heating profile is updated at every time-step as well. A number of studies in the past have concentrated all of the (tidal) heating in the deep interior of the planet \citep{2001ApJ...548..466B, 2003ApJ...592..555B, 2010ApJ...713..751I}. It is not known whether the tidal heating is deep-seated; it may be in the atmosphere and thus unavailable for inflation. By contrast the Ohmic model explicitly computes the relative amounts in the interior and atmosphere, so the actual calculated heating distribution is implemented in the structural calculations. Note that in our model, heating of the adiabatic interior as well as heating of the deep atmosphere (as in  \cite{2002A&A...385..156G}) contribute to the radial evolution. 

We solve the simplified steady-state induction equation (see \cite{2010ApJ...714L.238B} for derivation)
\begin{equation}
\nabla \cdot \sigma \nabla \Phi = \nabla \cdot \sigma \left( \vec{v} \times \vec{B}_{dip} \right)
\end{equation}
All of these planets can be expected to have internal dynamos with field strengths similar to Jupiter ($|\vec{B}| \sim $ few Gauss) or perhaps larger because of the lower density in the dynamo generating region. It is implicitly assumed the magnetic field remains roughly dipolar at the planetary surface, in spite of the induced current. This assumption may be generally valid in most hot Jupiter atmospheres, as the magnetic Reynold's number, $Re_m  = V L / \eta$, where $\eta$ is the magnetic diffusivity, is typically of order unity at most, rendering dynamo action unlikely in the atmosphere. However, it is unclear if the induced current from the atmosphere will have an appreciable effect on the interior dynamo, and its role should be assessed with a detailed, numerical model. As in \citep{2010ApJ...714L.238B}, we used a simple analytical prescription for the velocity field and the flow is confined to the radiative atmosphere so that that the RHS of the above equation vanishes in the (assumed) rigidly rotating interior. Nearly solid body rotation of the deep interior is predicted by the same argument that was used to reach this conclusion for Jupiter \citep{2008Icar..196..653L}. We assume that this is also the tidally synchronized state \citep{1963MNRAS.126..257G,1981A&A....99..126H}. Note that the assumption of tidal synchronization is present both, explicitly in our calculation of $\vec{v} \times \vec{B}_{dip}$, and implicitly in interpretation of GCMs of tidally locked planets. Once the electric scalar potential is known, the current, $J$, is determined from Ohm's law (including the induction emf i.e. equation 3). Thus, the Ohmic dissipation per unit mass, $\mathbb{P} = J^2/ \rho \sigma$ everywhere in the planet can also be computed.

\begin{figure}[t]
\includegraphics[width=0.5\textwidth]{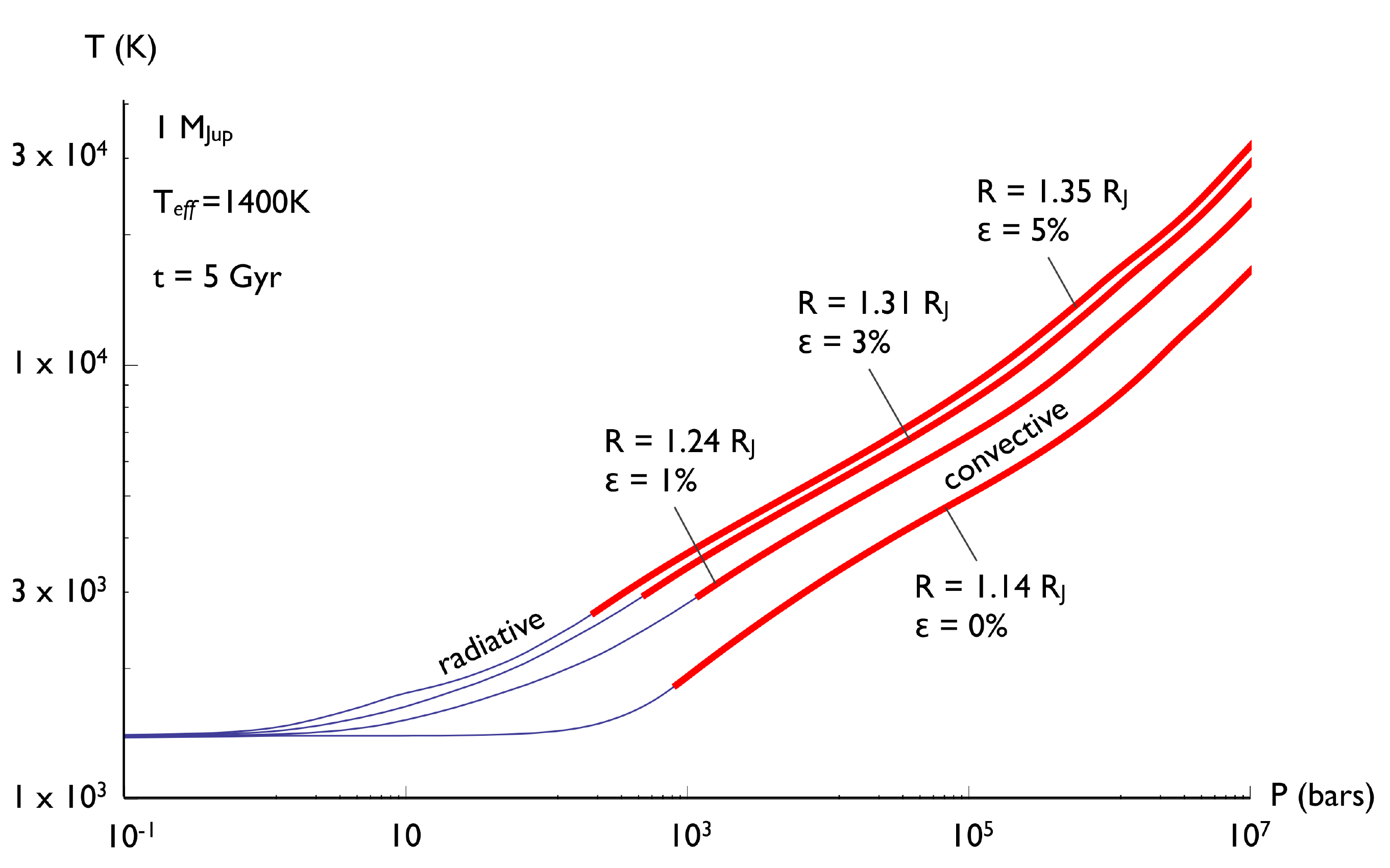}
\caption{A series of representative pressure-temperature profiles of an evolved ($t=4.5$Gyr) $1 M_{Jup}$ planet, with $T_{eff}=1400K$. The convective parts of the planet are plotted as thick red lines, while the radiative parts is plotted as thin purple lines. The four profiles correspond to $\epsilon = 0\%,1\%,3\%$ and $5\%$ solutions. Note: the published ApJ version of this figure contains a plotting error.} 
\end{figure}

As already discussed above, we can interpret the qualitative characteristics of the velocity field from GCM results. Although the details of the flows that GCMs produce vary somewhat, depending on the assumptions implicit to the solver \citep{2011MNRAS.tmp..370H}, most solutions produce a broad single jet in the atmosphere. The physical mechanism for the jet's formation has been identified as standing planetary-scale Rossby and Kelvin waves \citep{2011arXiv1103.3101S}. However, there exists a notable exception in the literature, where retrograde jets are also produced \citep{2010ApJ...716..144T}. This may be of importance for the Ohmic mechanism, since the atmospheric flows govern the induction of the interior current. To explore the effect of retrograde motions, we have computed the interior currents in a static model (as in \citet{2010ApJ...714L.238B}) with three different prescriptions for the jets. Namely, we considered, a single broad jet ($v_{\phi} \propto \sin \theta $), two counter-rotating jets, one in each hemisphere ($v_{\phi} \propto \sin 2\theta$), as well as three jets, with retrograde equatorial flow ($v_{\phi} \propto \sin3\theta$). 

The results of these calculations are presented in side-view cross-sections of planets in Figure 1. In each of the three panels, the small arrows inside the planet show the induced current (actual model output). Additionally, the magnetic field lines as well as graphical representations of the currents and atmospheric flows are presented. Figure 1A shows the current induced by a single jet. As in the results of \citet{2010ApJ...714L.238B}, loops are set up such that the radial current, which is induced at the equator flows towards the poles in the interior, and turns around in the atmosphere. If the current is induced by a double jet (Figure 1B), its geometry changes considerably. Since the flows produce counter-acting electro-motive forces, the interior current now flows from one hemisphere to the other. However, the hemispheric symmetry is restored with three jets (Figure 1C), and due to a particular choice of signs (namely, prograde flow at the equator in Figure 1A and retrograde flow at the equator in Figure 1C), the direction of the interior current is reversed. 

\begin{figure}[t]
\includegraphics[width=0.49\textwidth]{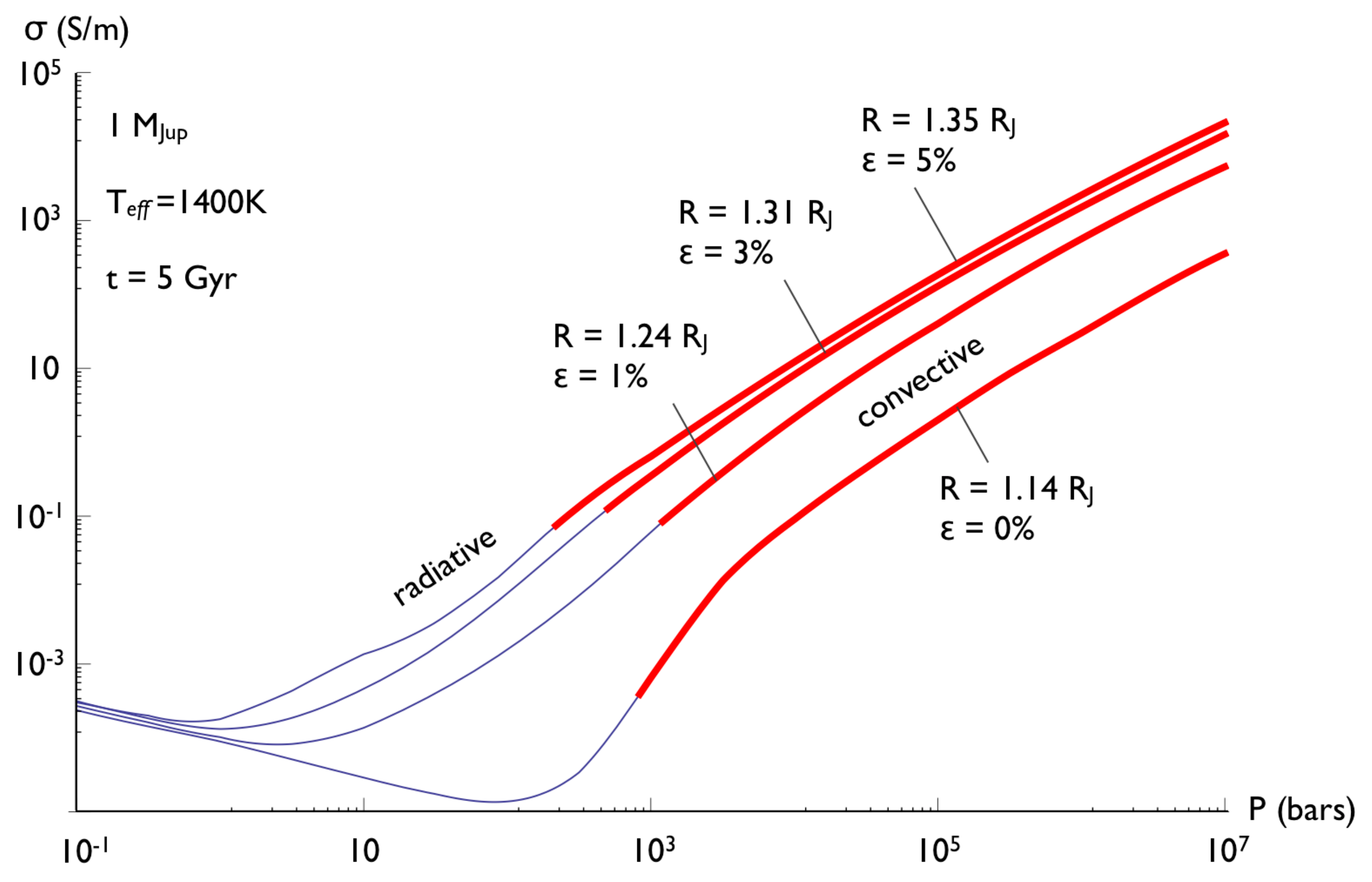}
\caption{A series of representative electrical conductivity profiles of an evolved ($t=4.5$Gyr) $1 M_{Jup}$ planet, with $T_{eff}=1400K$. The ionization was derived from the the temperature-pressure profiles, presented in Figure 2. The convective parts of the planet are plotted as thick red lines, while the radiative parts is plotted as thin purple lines. The four profiles correspond to $\epsilon = 0\%,1\%,3\%$ and $5\%$ solutions. Note: the published ApJ version of this figure contains a plotting error.} 
\end{figure}

Although, the interior current geometry changes considerably depending on the assumed flow, the resulting heating rate does not. Indeed, a comparison of the Ohmic dissipation profiles in the three static models described above reveals that the ratio of the cumulative dissipation below the radiative-convective boundary to the total is the same to within $\sim 10 \%$. Consequently, in accord with most GCM results, we approximate the atmospheric flow as a single zonal jet in all following calculations.

Our calculations of planetary interior structure and evolution employed a descendant of the Berkeley stellar evolution code \citep{1964ApJ...139..306H}. Over more than 40 years of its lifetime, the model has been subject to  a considerable amount of modification and improvement in its input physics and has been used extensively in the study of both extra-solar and solar system giant planets \citep{1996Icar..124...62P, 2003ApJ...592..555B, 2009ApJ...704L..49B}. A given evolutionary sequence starts at a radius of roughly $2 R_{Jup}$ and ends at an age of 5 Gyr.

The program we used to compute the structural evolution of hot Jupiters assumes that the standard Lagrangian equations of stellar structure apply. Energy transport is accomplished either by convection or radiation, as dictated by the Schwarzschild criterion. Energy sources within the planet include gravitational contraction, cooling of the interior, and Ohmic dissipation. For the gaseous envelope, the interpolated \cite{1995ApJS...99..713S} equation of state is used.  In the models with a core, the solid/liquid core has constant density. The imperfections of the equation of state are expected to have a much smaller effect on the computed radii than the uncertainties in our model for Ohmic inflation. 

The atmosphere is taken to be gray. Pure molecular opacities are used in the radiative outer layers of the planet \citep{2008ApJS..174..504F} while the high temperature and pressure, opacities of \cite{1994ApJ...437..879A} are used in the interior. In the atmosphere, it is assumed that dust grains contribute negligibly to the opacity. This may not be true but is again a smaller uncertainty than the other uncertainties in the modeling. We refer the reader to \cite{2005Icar..179..415H} and \cite{2008ApJ...688L..99D} for further reading. Given the approximate treatment of magnetic induction in the model, it is sensible to utilize a simple gray atmosphere. However, it is noteworthy that a different atmospheric model could modify our results on a quantitative level, since the atmospheric boundary condition directly governs the cooling rate of the interior \citep{2011A&A...527A..20G}. 

A series of representative temperature-pressure profiles of an evolved $1 M_{Jup}$ planet are plotted in Figure 2. The four presented evolutionary sequences were started from the same initial condition, but were evolved with different amount of Ohmic heating, namely $\epsilon = 0$, $\epsilon = 1\%$, $\epsilon = 3\%$ and $\epsilon = 5\%$, as labeled. It is noteworthy that in planets with higher $T_{eff}$, Ohmic dissipation in the atmosphere can lead to a formation of a small convective region, in the otherwise radiative atmosphere. Electrical conductivity profiles, corresponding to the presented temperature-pressure profiles are shown in Figure 3.

\begin{figure}[t]
\includegraphics[width=0.5\textwidth]{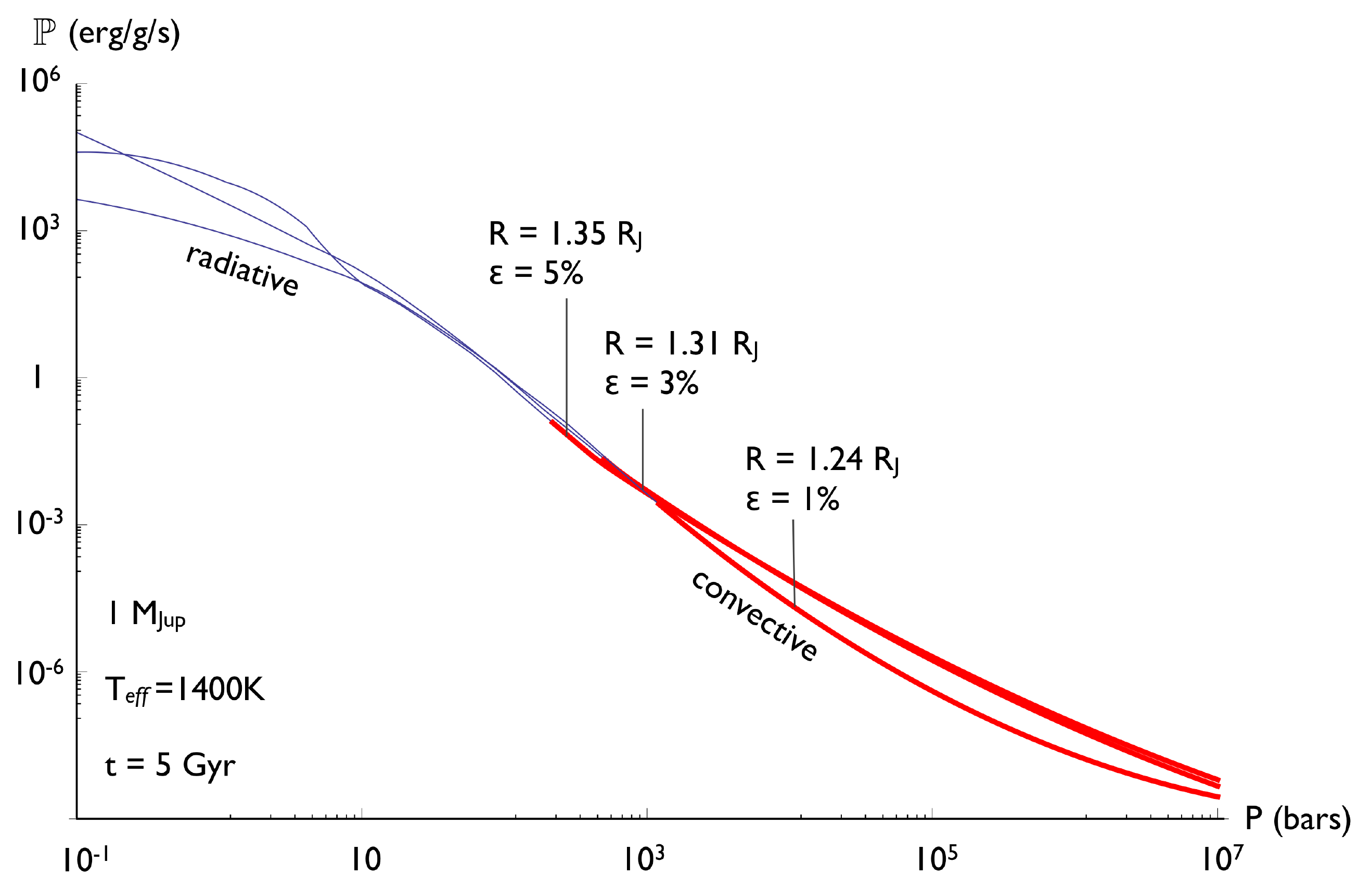}
\caption{A series of representative Ohmic heating profiles (energy dissipation rate per unit mass) of an evolved ($t=4.5$Gyr) $1 M_{Jup}$ planet, with $T_{eff}=1400K$, corresponding to the $\epsilon \neq 0$ temperature-pressure profiles shown in Figure 2. The thick red lines denote the heating of the adiabatic interior, while the thin purple lines represent the heating of the radiative atmospheres. Note: the published ApJ version of this figure contains a plotting error.} 
\end{figure}

At the planetary surface, defined as the Rosseland mean photosphere, the luminosity is composed of two parts: the internal luminosity generated by the planet, and the energy absorbed from the stellar radiation flux and re-radiated (insolation). In our simulations, the atmospheric circulation spans the region between $10$ bars and the photosphere, in accord with GCM results \citep{2009ApJ...699..564S}. Additionally, in the heating calculation, a boundary condition of zero radial electrical current is employed at the surface, as a result of significant decrease in conductivity with height in the upper atmosphere \citep{2010ApJ...714L.238B}. This decrease in electrical conductivity is not explicitly present in our models, since the temperature at the photosphere is $T_{eff}$ by definition. However, significant drops in temperatures shortly above the photosphere can be observed in more sophisticated atmospheric models (e.g. \cite{2009ApJ...699.1487S}). Furthermore, if a given atmosphere exhibits a thermal inversion, the characteristic decrease in temperature at $P \sim 10^{-2}$ bars provides a thin, electrically insulating layer. In other words, the null radial current boundary condition at the photosphere is likely to hold true for most atmospheres, whether or not an inversion exists. 

A series of representative Ohmic heating profiles (corresponding to the $\epsilon \neq 0$ temperature-pressure profiles presented in Figure 2) are shown in Figure 4. Note that although the heating is maximum at the very surface, the layers that contribute to the expansion lie much deeper (i.e. below $\sim 10$ bars at an early epoch and below $\sim 10^2 - 10^3$ bars for evolved planets). The location of maximal heating corresponds to the atmospheric height where the electrical current turns around. Thus the higher up, electrically insulating layers of the atmosphere discussed in the previous paragraph, would experience very little Ohmic heating.

\section{Results: Radial Evolution}
We computed the structural evolution of Ohmically heated hot Jupiters with the mass range spanning an order of magnitude between $0.23 M_{Jup}$ and $3 M_{Jup}$ and effective temperature range between $1000K$ and $1800K$. All but one of our models are core-less, and thus should be viewed as giving an upper bound on the radius that a planet of a given mass and $T_{eff}$ may achieve. In our suite of simulations, we observed two distinct families of solutions that are characterized by either a monotonically decreasing or a monotonically increasing radius in time. A few representative evolution sequences are presented in Figure 5, with their masses and effective temperatures labeled. In the solutions where the radius is monotonically decreasing with time, the final answer (radius at $t=5$Gyr) is largely independent of the initial condition: the solution always asymptotically approaches the same ``equilibrium radius." If the radius monotonically increases with time, however, the quantity of importance is the total integrated heating rate, which is intrinsically a function of the planetary radius. Consequently, these ``unstable" solutions are unavoidably dependent on the initial conditions.

\begin{figure}[t]
\includegraphics[width=0.5\textwidth]{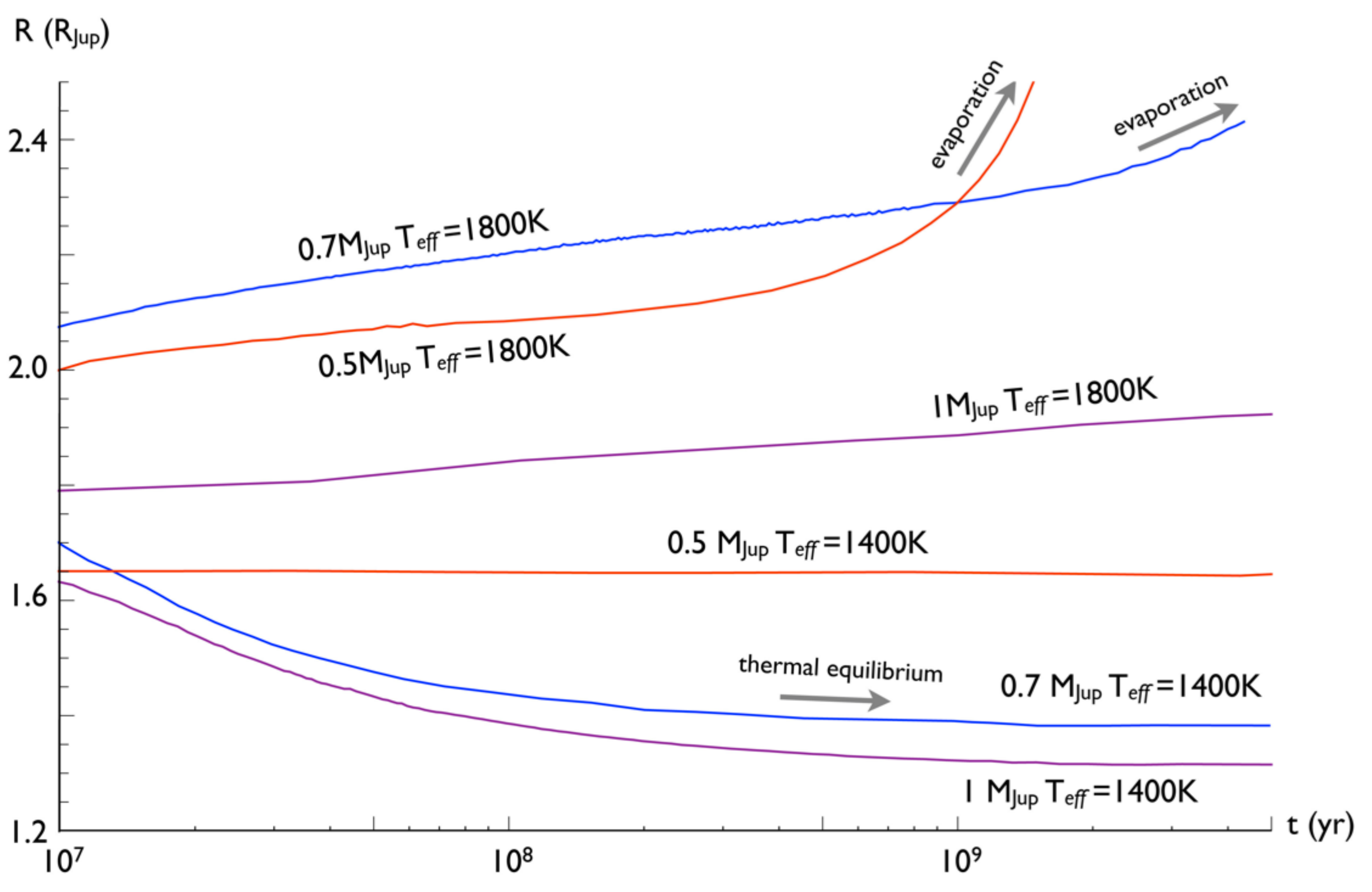}
\caption{A sample of evolution sequences of $0.5, 0.7$ and $1 M_{Jup}$} coreless planets at $T_{eff} = 1400$ and $1800K$ with $\epsilon = 3\%$. The $0.7$ and $1 M_{Jup}$ planets at $1400K$ planets asymptotically approach radii that are characterized by thermal equilibrium. The $0.5, 0.7$ and $1 M_{Jup}$ planets at $1800K$ are all on unstable evolutionary paths which eventually lead to Roche-Lobe overflow. However, the $1 M_{Jup}$ planet, however, is considerably more stable, owing to its more degenerate state. The evolution of the $0.5 M_{Jup}$ planet at $1400K$ lies between the two regimes - its starting radius is the equilibrium radius.
\end{figure}

Hot Jupiters can change their radii significantly primarily because of a  change in the entropy of the deep interior, since the outermost several (radiative) scale heights constitute only a small fraction of the planet radius. In the idealized limit where the heat escaping from the deep interior is small  compared to Ohmic heating, $\mathcal{P}$, the equation governing evolution for the deep interior reads
\begin{equation}
\frac{d}{dt}(E_{\textrm{grav}}+E_{\textrm{int}}) = \mathcal{P}.
\end{equation}
If the planet is a polytrope and the internal energy is entirely thermal then the sum of internal and gravitational energies can be written as $- k GM^2/R$ where $k$ is a constant typically somewhat smaller than unity \citep{1957QB461.C45......}.  In the case of highly degenerate bodies (e.g. Jupiter itself) the zero temperature part of the internal energy changes in such a way as to exactly cancel the change in gravitational energy as radius changes \citep{1984plin.book.....H} and the LHS above becomes only the time derivative of thermal energy to a good approximation. (This invalidates the claim, often made but erroneous, that the luminosity of degenerate planets is derived from contraction). The bodies of interest to us have non-ideal thermodynamics and 
are not in the degenerate limit so no simple result holds. However, direct calculation shows that it is still approximately true to replace the LHS above with something like $d/dt(-kGM^2/R)$ with $k$ of order unity, even though it is not correct to think of the energy as being derived solely from changes in gravitational energy. All of this discussion ignores possible effects arising from redistribution of the heavy elements. 

The transition from stable to unstable solutions can only be understood through detailed models, because it depends on the details of opacity and conductivity structure, but the essential physics lies in the comparison of Ohmic heating at depth with the total radiative heat loss from the convective interior evaluated at the radiative/convective boundary. This heat loss scales as the adiabatic temparature gradient, which depends in turn on $g$, the gravitational acceleration. This is approxcimately linear in mass because radius does not vary much, so Ohmic heating overwhelms heat loss from the interior once the mass is sufficiently small. In the limit where the Ohmic heating dominates the evolution of the deep interior, we expect that the 
characteristic timescale of inflation $\tau_{infl} \equiv R/(dR/dt)$ is (to order of magnitude) the ratio of $GM^2/R$ to Ohmic heating, or 
\begin{equation}
\tau_{\textit{infl}}\sim (\frac{0.01}{\epsilon})(\frac{M}{M_{J}})^2 (\frac{R_{J}}{R})^3 (\frac{1500 \textrm{K}}{T_{eff}})^4 \textrm{Gyr}.
\end{equation}

As can be seen in Figure 5, after a sufficient amount of time, unstable solutions enter a phase of runaway growth, with the instability driven by the fact that the cumulative heating rate is proportional to the surface area of the planet. This inevitably leads to Roche-lobe overflow and evaporation of the planet \citep{2008ApJ...685..521L}. The timescale over which the planet can remain intact while on an unstable path is a function of the planetary mass and high-mass planets can, in principle, remain intact and grow slowly over many Gyr, while low-mass giant planets evaporate on a $\sim 1$Gyr timescale. It is noteworthy that the timescale will also likely be affected by the $1/R$ slowdown of winds (see equation 16) and thus a possible modification of the efficiency (which we kept constant) with a growing radius. Still, it is very likely that evaporation is unavoidable in certain circumstances. 

 \begin{figure}[t]
\includegraphics[width=0.5\textwidth]{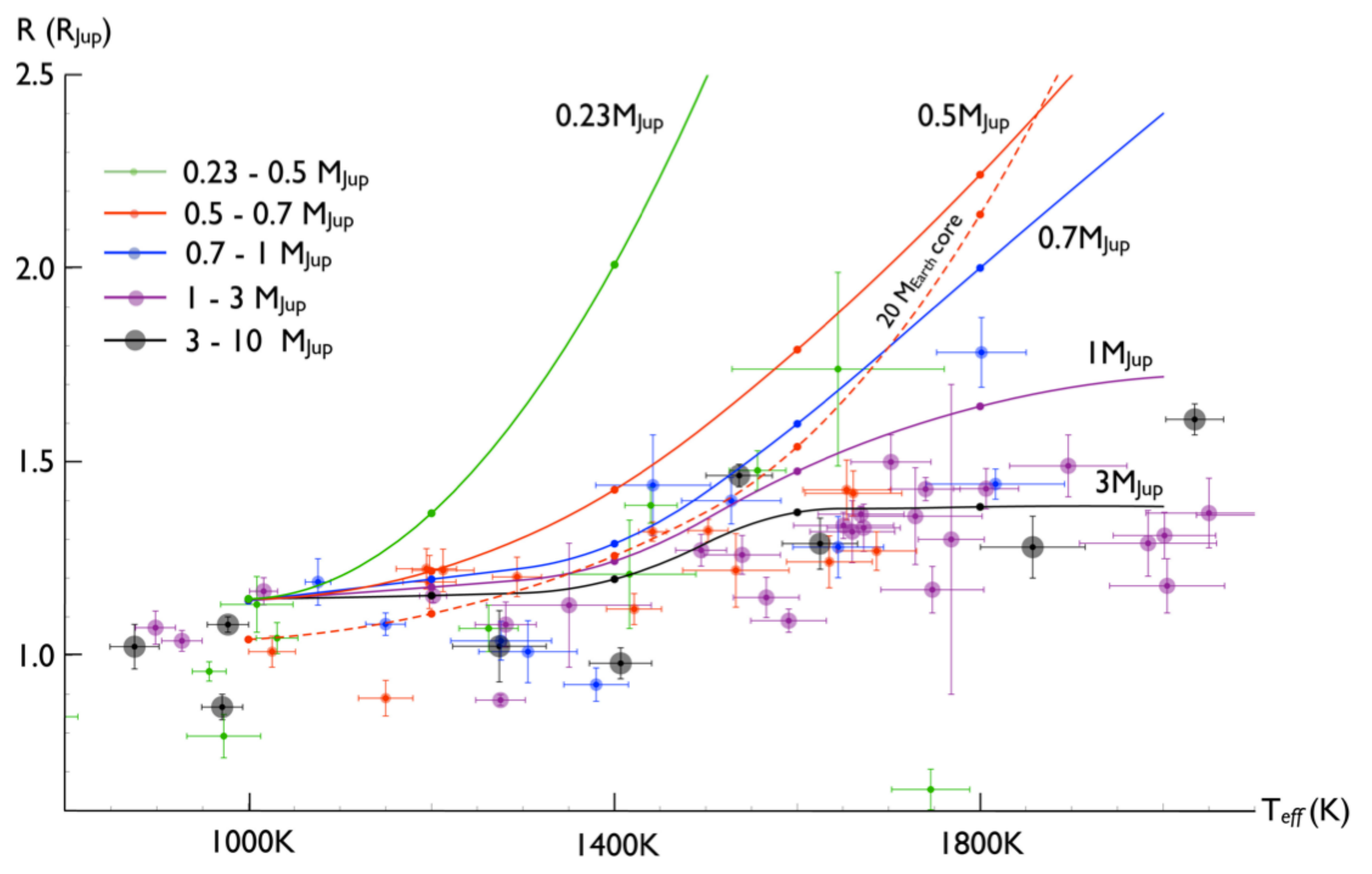}
\caption{An $R-T_{eff}$ diagram of evolved models of various masses ($\epsilon = 1\%$) as well as the current aggregate of detected transiting hot Jupiters.  The solid lines represent theoretical curves, corresponding to core-less models, while the dashed line depicts $M = 0.5M_{Jup}$ models with a $20 M_{\oplus}$ solid core. The observational data points are presented, along with their corresponding $1 \sigma$ error bars. The masses of the data are binned to fall between the theoretical curves and are labeled by color and size as shown on the figure.} 
\end{figure}

Low-mass planets can be stabilized against evaporation, if they possess sufficiently massive high-metallicity cores \citep{2003ApJ...592..555B, 2010ApJ...713..751I}. However the critical core mass is far from being trivially small. In fact, a simulation of an $0.5M_{Jup}$ planet with a $\rho = 3$g/cc, $20 M_{\oplus}$, core at $T_{eff}=1800K$ reveals that it is only able to retain its envelope for $\sim 1$ Gyr before overflowing the Roche-lobe. Given that hot low-mass planets tend to reside in compact multiple systems \citep{2007ApJ...654.1110T}, while hot Jupiters usually lack close-by companions \citep{2010A&A...512A..48L}, the discussion of evaporation is suggestive of the possibility that a significant fraction of hot, sub-giant planets, without companions, may have been born as giant planets and have since lost their gaseous envelopes\footnote{A similar idea has already been proposed in the context of thermal atmospheric escape \citep{2005A&A...436L..47B}.}.

Within the context of our mechanism, the planetary radius is most clearly visualized as function of effective temperature rather than mass. Consequently, we choose to plot equal-efficiency, equal-mass lines on an $R-T_{eff}$ diagram rather than a conventional $R-M$ diagram. Our results, along with the current aggregate of well-characterized transiting hot Jupiters, are illustrated in Figures 6-8. The evolved final model radii are plotted as points with spline interpolation between them and are labeled according to their masses. The observational data\footnote{The data was acquired from the Extrasolar Encyclopedia (http://www.exoplanet.eu)} are plotted as points with their corresponding $1 \sigma$ error bars. The data are binned in mass, to fall between our model masses, and are labeled by color. As a result, if the Ohmic mechanism is indeed correct, and unmodeled effects such as variable opacity and age make a small contribution \citep{2007ApJ...661..502B}, all observed data of a given color should fall below the model curve of the same color.

\begin{figure}[t]
\includegraphics[width=0.5\textwidth]{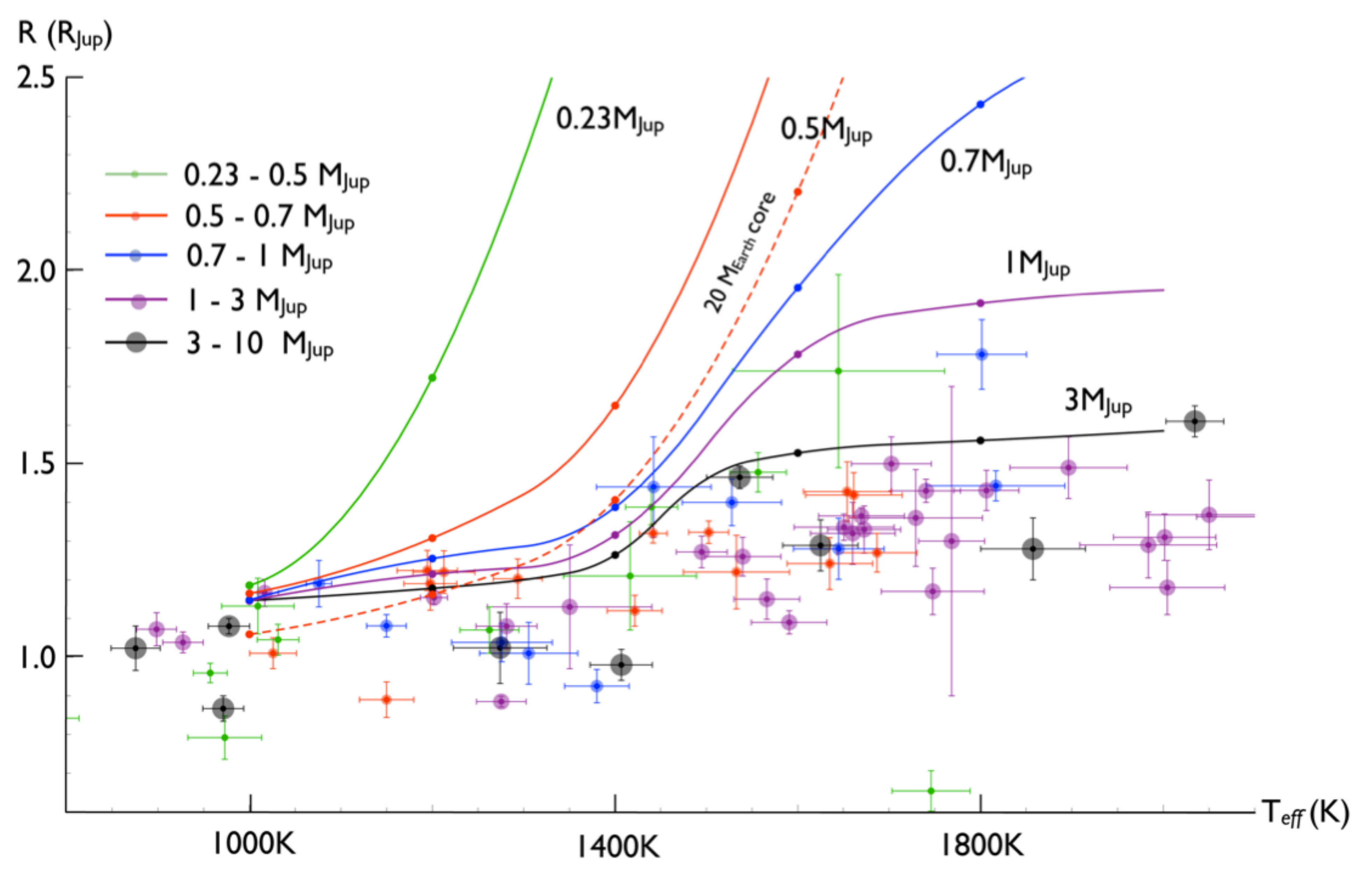}
\caption{Same as Figure 2, but with $\epsilon = 3\%$} 
\end{figure}

\section{Discussion}

As is clearly demonstrated, if the approximations inherent to our model are valid, Ohmic dissipation, along with the likely presence of high-metalicity cores, is able to account for all of the currently observed radius anomalies. Furthermore, the self-similar family of model curves reproduces the characteristic increase and overturn of the radii in the effective temperature region between $1200K$ and $1800K$. This feature is unique to the Ohmic dissipation mechanism and is a consequence of ionization. Recall that the primary source of electrons in the atmosphere is  alkali metals,  with K playing a dominant role. Under densities that are characteristic of hot Jupiter atmospheres, K has an ionization temperature around $\sim 1800K$. This explains the rapid increase in Ohmic heating with rising $T_{eff}$ around $\sim 1400$K. This increase comes about through a Boltzmann factor in the Saha equation, and will therefore have a roughly exponential dependence. We stress that this is uniquely characteristic of the Ohmic mechanism, since other proposed mechanisms will exhibit an algebraic dependence on $T_{eff}$ (see \cite{2011ApJL...Laughlin} for a statistical comparison of various inflation mechanisms with data).

As the temperature nears $\sim 1800$K,  the ionization levels begin to saturate and the radiative/convective boundary moves deeper, so the radii don't grow as quickly with  $T_{eff}$. However ionization of Na begins to contribute strongly above $2000K$. This renders all but the most massive of our models unstable. This is suggestive, given that out of a sample of 87 transiting Hot Jupiters, only four have been detected with $T_{eff} > 2100K$, although the mechanism by which Hot Jupiters halt their migration at small radii is still not fully understood.

It is noteworthy that our low-mass models significantly overestimate the data, to the extent where small ($\sim 20 M_{\oplus}$) cores are insufficient to explain the observed radii. This is not particularly surprising, given that the inflation is somewhat enhanced by a shallower adiabatic temperature gradient, $g/c_p$, allowing for a larger fraction of the generated heat to be deposited into the convective interior, and a diminished gravitational energy, that must be overcome by Ohmic heating. The discrepancy between the model and the data is either indicative of massive cores, or more likely curbed efficiency of the Ohmic mechanism at smaller mass. Both are certainly plausible, and the latter can be a consequence of diminished strength of the magnetic field, which is dictated by the intrinsic heat-flux \citep{2009Natur.457..167C}.  

\begin{figure}[t]
\includegraphics[width=0.5\textwidth]{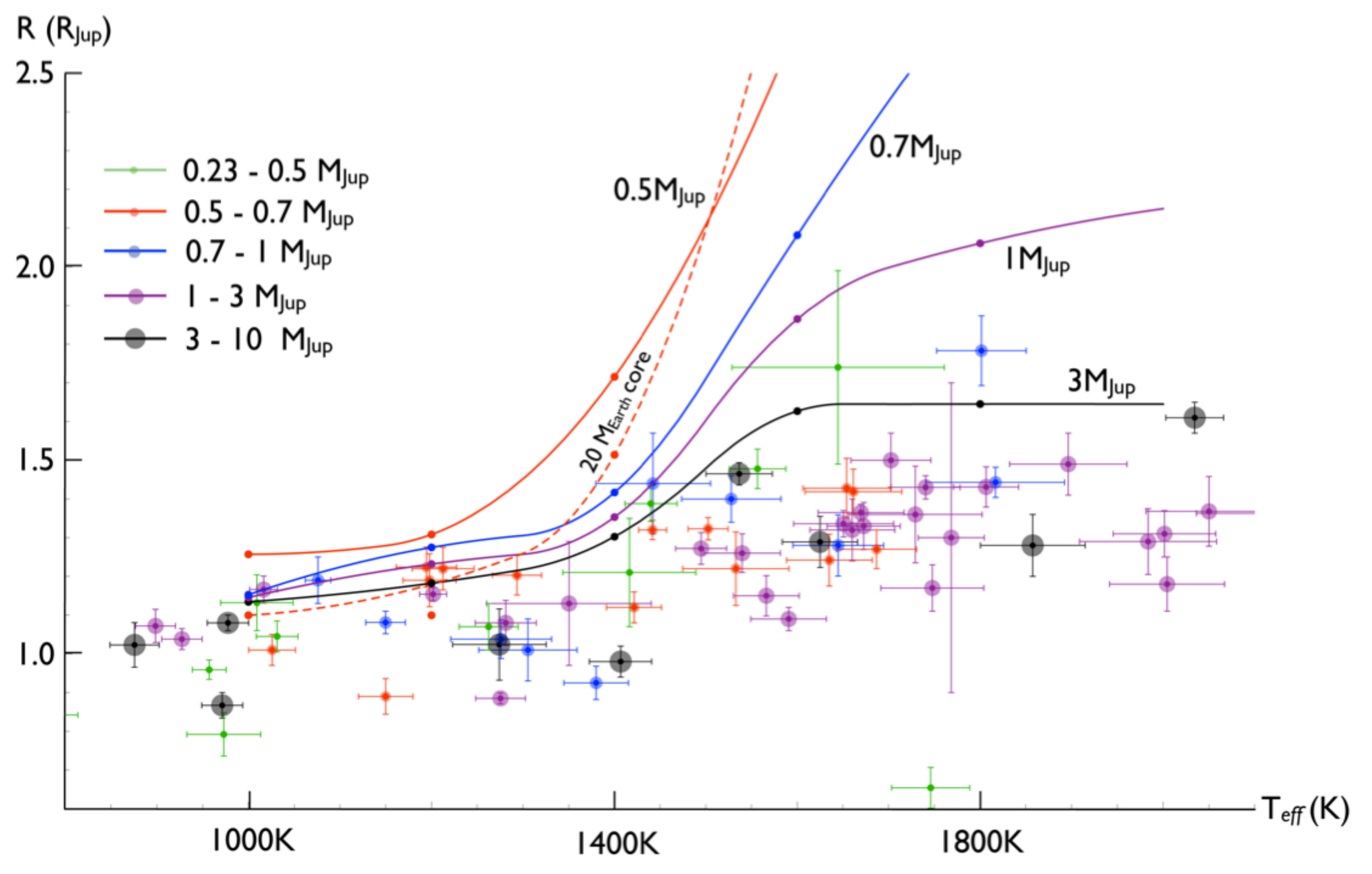}
\caption{Same as Figure 2, but with $\epsilon = 5\%$} 
\end{figure}

As the parameter space available to hot Jupiters is, to first order, delineated, the models presented in this work make two predictions. First and foremost, the radii of all hot Jupiters should fall below the theoretical curves presented in Figures 6-8. Second, a fraction of the hot sub-giant planet population may have originated from the giant planet population. Since this process can take place many Gyr after the birth of the star, Ohmically evaporating planets should be detected, given a sufficiantly large data set. These predictions will become testable, as the known aggregate of transiting exoplanets continues to grow. Accordingly, additional data will allow us to further understand the diversity of planetary bodies and  to better define the Solar System's place in our galactic neighborhood. 

We would like to conclude with a discussion of future prospects for improvement of the model. Let us begin by recalling the simplifications inherent to the approach presented here. In the structural calculation, we used a gray, one-dimensional atmosphere. Obviously this is not intended to be a quantitatively good representation of real atmospheres and improved treatment of the temperature (and thus conductivity) profile, accounting for asymmetry in insolation between the day and the night side as well as heat transport by global circulation is desired in the future. Indeed the considerable difference in interior cooling rates shown here, and those presented by \citet{2011A&A...527A..20G} suggests that the effect of changing the atmospheric boundary condition on structural evolution may be considerable. That said, the inability of current state-of-the-art atmospheric models to match observations in a satisfactory manner suggests that much additional effort must be applied towards improvement of our understanding of extra-solar planet atmospheres for implementation of a complex atmosphere into our model to be meaningful. In the heating calculation, our analytical parameterization of the circulation should be replaced by a self-consistent solution of the global circulation that fully accounts for magneto-hydrodynamic effects. The induced current in the interior would be a natural result of such a calculation. At the same time, the effect of the induced current on the interior dynamo should be assessed. Although in this study we assumed that the interior dynamo is largely unaffected by the currents induced in the atmosphere, in reality this may not be the case, since the induced current can in principle be comparable to the characteristic current of the interior dynamo. However, given the enormous computational cost of such an endeavor, it is not surprising that a calculation of this sort has not yet been done. 

In general, future efforts in improvement of the Ohmic heating model should rely on both, modeling and observational constraints. After all, observations of primary and secondary eclipses can often yield information about the atmospheric temperature profile \citep{2008ApJ...673..526K} as well as elemental abundances  \citep{2010ApJ...717..496L}. At the same time, ground-based high-dispersion spectroscopy can yield an estimate of the wind velocity \citep{2010Natur.465.1049S}. In certain cases (e.g. WASP-12b), the planet's proximity to its host star results in considerable interactions between the planetary magnetosphere and the stellar wind. This can lead to formation of a bow-shock, which ultimately produces an observable signature, that manifests itself as an early ingress of the transit in near-UV frequency range. Quantitative modeling of this process may lead to meaningful constraints on the planetary magnetic field \citep{2010ApJ...722L.168V}. As a result, an observationally guided improvement of the Ohmic inflation model will surely prove to be a useful venture, as quantitative enhancement of the first-order results presented in this paper will allow for a rigorous statistical comparison between the model and an expanded data set.
\\
\\
\textbf{Acknowledgments} We are indebted to A. Ingersoll, G. Laughlin, T. Guillot and T. Schneider for useful conversations, as well as the referee, Jonathan Mitchell for providing helpful suggestions. Additional thanks to Andrew Cumming for alerting us to the graphing errors in Figures 2-4.


\begin{thebibliography} 

\bibitem[Alexander 
\& Ferguson(1994)]{1994ApJ...437..879A} Alexander, D.~R., \& Ferguson, J.~W.\ 1994, \apj, 437, 879

\bibitem[Batygin 
\& Stevenson(2010)]{2010ApJ...714L.238B} Batygin, K., \& Stevenson, D.~J.\ 2010, \apjl, 714, L238 

\bibitem[Batygin et al.(2009)]{2009ApJ...704L..49B} Batygin, K., 
Bodenheimer, P., \& Laughlin, G.\ 2009, \apjl, 704, L49 

\bibitem[Baraffe et al.(2005)]{2005A&A...436L..47B} Baraffe, I., Chabrier, G., Barman, T.~S., Selsis, F., Allard, F., \& Hauschildt, P.~H.\ 2005, \aap, 436, L47

\bibitem[Bodenheimer et al.(2001)]{2001ApJ...548..466B} Bodenheimer, P., 
Lin, D.~N.~C., \& Mardling, R.~A.\ 2001, \apj, 548, 466 

\bibitem[Bodenheimer et al.(2003)]{2003ApJ...592..555B} Bodenheimer, P., 
Laughlin, G., \& Lin, D.~N.~C.\ 2003, \apj, 592, 555 

\bibitem[Burrows et al.(2007)]{2007ApJ...661..502B} Burrows, A., Hubeny, 
I., Budaj, J., \& Hubbard, W.~B.\ 2007, \apj, 661, 502 

\bibitem[Chabrier 
\& Baraffe(2007)]{2007ApJ...661L..81C} Chabrier, G., \& Baraffe, I.\ 2007, \apjl, 661, L81

\bibitem[Chandrasekhar(1957)]{1957QB461.C45......} Chandrasekhar, S.\ 1957, 
[New York] Dover Publications [1957],  

\bibitem[Cho et al.(2008)]{2008ApJ...675..817C} Cho, J.~Y.-K., Menou, K., 
Hansen, B.~M.~S., \& Seager, S.\ 2008, \apj, 675, 817 

\bibitem[Christensen et al.(2009)]{2009Natur.457..167C} Christensen, U.~R., 
Holzwarth, V., \& Reiners, A.\ 2009, \nat, 457, 167 

\bibitem[Dodson-Robinson et al.(2008)]{2008ApJ...688L..99D} 
Dodson-Robinson, S.~E., Bodenheimer, P., Laughlin, G., Willacy, K., Turner, 
N.~J., \& Beichman, C.~A.\ 2008, \apjl, 688, L99 

\bibitem[Fortney 
\& Nettelmann(2010)]{2010SSRv..152..423F} Fortney, J.~J., \& Nettelmann, N.\ 2010, \ssr, 152, 423

\bibitem[Freedman et al.(2008)]{2008ApJS..174..504F} Freedman, R.~S., 
Marley, M.~S., \& Lodders, K.\ 2008, \apjs, 174, 504 

\bibitem[Guillot(1999)]{1999Sci...286...72G} Guillot, T.\ 1999, Science, 
286, 72 

\bibitem[Guillot \& Showman(2002)]{2002A&A...385..156G} Guillot, T., \& Showman, A.~P.\ 2002, \aap, 385, 156

\bibitem[Guillot \& Havel(2011)]{2011A&A...527A..20G} Guillot, T., \& Havel, M.\ 2011, \aap, 527, A20 

\bibitem[Goldreich(1963)]{1963MNRAS.126..257G} Goldreich, P.\ 1963, \mnras, 
126, 257 

\bibitem[Heng et al.(2011)]{2011MNRAS.tmp..370H} Heng, K., Menou, K., 
\& Phillipps, P.~J.\ 2011, \mnras, 370 

\bibitem[Henyey et al.(1964)]{1964ApJ...139..306H} Henyey, L.~G., Forbes, 
J.~E., \& Gould, N.~L.\ 1964, \apj, 139, 306 

\bibitem[Held \& Hou(1980)]{1980JAtS...37..515H} Held, I.~M., \& Hou, A.~Y.\ 1980, Journal of Atmospheric Sciences, 37, 515 

\bibitem[Holton(1992)]{1992aitd.book.....H} Holton, J.~R.\ 1992, 
International geophysics series, San Diego, New York: Academic Press, 
|c1992, 3rd ed.,  

\bibitem[Hubbard(1984)]{1984plin.book.....H} Hubbard, W.~B.\ 1984, New 
York, Van Nostrand Reinhold Co., 1984, 343 p., 

\bibitem[Hubickyj et al.(2005)]{2005Icar..179..415H} Hubickyj, O., 
Bodenheimer, P., \& Lissauer, J.~J.\ 2005, Icarus, 179, 415 

\bibitem[Hut(1981)]{1981A&A....99..126H} Hut, P.\ 1981, \aap, 99, 126 

\bibitem[Ibgui et al.(2010)]{2010ApJ...713..751I} Ibgui, L., Burrows, A., 
\& Spiegel, D.~S.\ 2010, \apj, 713, 751 

\bibitem[Knutson et al.(2008)]{2008ApJ...673..526K} Knutson, H.~A., 
Charbonneau, D., Allen, L.~E., Burrows, A., 
\& Megeath, S.~T.\ 2008, \apj, 673, 526 

\bibitem[Laughlin et al.(2011)]{2011ApJL...Laughlin} Laughlin, G., Crismani, M., \& Adams, F. \  \apjl, submitted

\bibitem[Laine et al.(2008)]{2008ApJ...685..521L} Laine, R.~O., Lin, 
D.~N.~C., \& Dong, S.\ 2008, \apj, 685, 521 

\bibitem[Liu et al.(2008)]{2008Icar..196..653L} Liu, J., Goldreich, P.~M., \& Stevenson, D.~J.\ 2008, icarus, 196, 653 

\bibitem[Liu 
\& Schneider(2010)]{2010JAtS...67.3652L} Liu, J., \& Schneider, T.\ 2010, Journal of Atmospheric Sciences, 67, 3652

\bibitem[Lo Curto et 
al.(2010)]{2010A&A...512A..48L} Lo Curto, G., et al.\ 2010, \aap, 512, A48 

\bibitem[Line et al.(2010)]{2010ApJ...717..496L} Line, M.~R., Liang, M.~C., 
\& Yung, Y.~L.\ 2010, \apj, 717, 496 

\bibitem[Perna et al.(2010)]{2010ApJ...724..313P} Perna, R., Menou, K., 
\& Rauscher, E.\ 2010, \apj, 724, 313 

\bibitem[Peixoto 
\& Oort(1992)]{1992phcl.book.....P} Peixoto, J.~P., \& Oort, A.~H.\ 1992, New York: American Institute of Physics (AIP), 1992,  

\bibitem[Pollack et al.(1996)]{1996Icar..124...62P} Pollack, J.~B., 
Hubickyj, O., Bodenheimer, P., Lissauer, J.~J., Podolak, M., \& Greenzweig, Y. Icarus 1996, 124, 62 

\bibitem[Saumon et al.(1995)]{1995ApJS...99..713S} Saumon, D., Chabrier, 
G., \& van Horn, H.~M.\ 1995, \apjs, 99, 713

\bibitem[Schneider 
\& Lindzen(1977)]{1977JAtS...34..263S} Schneider, E.~K., \& Lindzen, R.~S.\ 1977, Journal of Atmospheric Sciences, 34, 263 

\bibitem[Showman et al.(2009)]{2009ApJ...699..564S} Showman, A.~P., 
Fortney, J.~J., Lian, Y., Marley, M.~S., Freedman, R.~S., Knutson, H.~A., 
\& Charbonneau, D.\ 2009, \apj, 699, 564 

\bibitem[Showman et al.(2010)]{2010exop.book..471S} Showman, A.~P., Cho, 
J.~Y.-K., \& Menou, K.\ 2010, Exoplanets, 471 

\bibitem[Showman and Polvani(2011)]{2011arXiv1103.3101S} Showman, A.~P., 
Polvani, L.~M.\ 2011.\ Equatorial superrotation on tidally locked 
exoplanets.\ ArXiv e-prints arXiv:1103.3101. 

\bibitem[Snellen et al.(2010)]{2010Natur.465.1049S} Snellen, I.~A.~G., de 
Kok, R.~J., de Mooij, E.~J.~W., \& Albrecht, S.\ 2010, \nat, 465, 1049 

\bibitem[Spiegel et al.(2009)]{2009ApJ...699.1487S} Spiegel, D.~S., 
Silverio, K., \& Burrows, A.\ 2009, \apj, 699, 1487 

\bibitem[Stevenson(1982)]{1982AREPS..10..257S} Stevenson, D.~J.\ 1982, 
Annual Review of Earth and Planetary Sciences, 10, 257 

\bibitem[Terquem 
\& Papaloizou(2007)]{2007ApJ...654.1110T} Terquem, C., \& Papaloizou, J.~C.~B.\ 2007, \apj, 654, 1110 

\bibitem[Thrastarson 
\& Cho(2010)]{2010ApJ...716..144T} Thrastarson, H.~T., \& Cho, J.~Y.\ 2010, \apj, 716, 144 

\bibitem[Vidotto et al.(2010)]{2010ApJ...722L.168V} Vidotto, A.~A., 
Jardine, M., \& Helling, C.\ 2010, \apjl, 722, L168

\bibitem[Youdin 
\& Mitchell(2010)]{2010ApJ...721.1113Y} Youdin, A.~N., \& Mitchell, J.~L.\ 2010, \apj, 721, 1113 

\end{thebibliography}
\end{document}